\documentclass[letterpaper]{article} 

\usepackage[super,comma,sort&compress]{natbib}
\usepackage{graphicx}
\usepackage{geometry}
\usepackage{siunitx}

\usepackage{authblk}

\title{High-quality Nano-patterning of Oxide Interfaces Using Transferred Gold Masks}

\author[1,2]{Qing Xiao}
\author[1]{Yanling Liu}
\author[1]{Changjian Ma}
\author[1]{Danqing Liu}
\author[1,3]{Zhiyuan Qin}
\author[1]{Qianyi Zhao}
\author[1,2]{Chengyuan Huang}
\author[1]{Mengke Ha}
\author[1]{Zhenhao Li}
\author[1,2,3]{Guanglei Cheng\thanks{Email: glcheng@ustc.edu.cn}}
\affil[1]
{Laboratory of Spin Magnetic Resonance, School of Physical Sciences, Anhui Province Key Laboratory of Scientific Instrument Development and Application, University of Science and Technology of China, Hefei 230026, China}
\affil[2]
{Hefei National Research Center for Physical Sciences at the Microscale, University of Science and Technology of China, Hefei 230026, China}
\affil[3]
{Hefei National Laboratory, University of Science and Technology of China, Hefei 230088, China}
\date{}

\begin{document}

\maketitle

\section*{Abstract}

Complex oxide interfaces, such as $\mathrm{SrTiO_3}$ and $\mathrm{KTaO_3}$ based heterostructures, host rich correlated phenomena with strong potential for advanced device applications. However, these interfaces are extremely susceptible to contamination and defect formation during nanofabrication, which often compromises device performance. Here, we present a solvent-free method for patterning oxide interfaces by employing high-resolution transferable thin metal masks in conjunction with oxygen-enriched $\mathrm{Ar^+}$ ion milling, which enables a clean and well-controlled nanofabrication process. Transport measurements demonstrate that the fabricated devices preserve their intrinsic properties, including high carrier mobilities, with negligible degradation compared to the pristine interfaces. This technique offers a convenient and robust route for engineering high-performance oxide electronic devices with precisely tailored transport characteristics.


\section{Introduction}

Oxide interfaces host a rich spectrum of emergent quantum phenomena, including two-dimensional electron gas (2DEG)\cite{ohtomo2004high}, metal-insulator transitions\cite{thiel2006tunable}, superconductivity\cite{reyren2007superconducting}, and spin-orbit coupling\cite{ben2010tuning,caviglia2010tunable}.
These properties arise from delicate electronic reconstruction and various symmetry-breaking states, and are prone to environmental perturbations due to the proximity of the interface and surface. To harness these effects for functional quantum devices, precise control over nanofabrication is essential, requiring methods that minimize structural and electronic disorder while enabling high-resolution patterning.

Current nanofabrication methods for oxide heterostructures can be broadly classified into two categories: top-down material removal\cite{paolo2013nano,d2021nanopatterning,minhas2016sidewall} and bottom-up selective deposition\cite{schneider2006microlithography,bell2009dominant,banerjee2012direct,stornaiuolo2012plane,gopinadhan2015gate,trier2015patterning,maniv2016tunneling,niu2017suppressed,fuchs2017patterning,bjorlig2018nanoscale}.
For perovskite substrates like $\mathrm{SrTiO_3}$ (STO) and $\mathrm{KTaO_3}$ (KTO), conventional $\mathrm{Ar^+}$ ion etching often induces undesirable conductive channels\cite{reagor2005highly,harashima2013coexistence}.
Modified methods including low energy\cite{paolo2013nano} or low temperature $\mathrm{Ar^+}$ ion etching\cite{d2021nanopatterning} and $\mathrm{BCl_3}$-based reactive ion etching\cite{minhas2016sidewall} have been explored to mitigate such parasitic conduction.
Alternatively, selective wet etching has proven potential for certain oxides\cite{bridoux2012alternative}.
In bottom-up approaches, oxides are epitaxially selectively deposited with predefined masks.
Conventional polymer masks decompose at elevated temperatures, restricting their use to room-temperature depositions\cite{bjorlig2018nanoscale}.
High-temperature compatible hard masks are widely used instead\cite{schneider2006microlithography,bell2009dominant,banerjee2012direct,stornaiuolo2012plane,gopinadhan2015gate,trier2015patterning,maniv2016tunneling,niu2017suppressed,fuchs2017patterning}.
However, these conventional approaches inevitably expose the critical interfaces or surfaces to polymer residues or chemical solutions, potentially inducing detrimental modifications to the interfacial electronic structure through unintended doping\cite{xie2011control,scheiderer2015surface}, defect formation, or surface redox reactions\cite{bristowe2011surface}.
While silicon stencil masks\cite{azimi2014nanoscale} can mitigate contamination, the fragility, complex fabrication process, relatively large size and edge blurring effect limit their applicability for complex device architectures.

Reconfigurable patterning has been realized in some oxide interfaces like $\mathrm{LaAlO_3}$/$\mathrm{SrTiO_3}$ (LAO/STO) and $\mathrm{LaAlO_3}$/$\mathrm{KTaO_3}$ (LAO/KTO) via conductive atomic force microscopy (cAFM)\cite{cen2009oxide} and focused electron beam irradiation\cite{yu2022nanoscale}, which control insulator-metal transitions locally and reversibly at extreme resolutions. Nanoscale structures created with high precision have exhibited exceptional quantum transport properties\cite{cheng2015electron,cheng2018shubnikov,briggeman2020pascal}. However, the stability of devices remains fundamentally limited by degradation in ambient conditions\cite{bi2010water}.

In this work, we present an advanced nanofabrication technique utilizing transferable Au films as hard masks in combination with oxygen-enriched $\mathrm{Ar^+}$ ion milling. The method successfully preserves the high-mobility characteristics of oxide interfaces after nanofabrication. Significantly, the fabricated devices achieved a feature size $\sim$\SI{100}{nm}, limited by the resolution of the Au masks we used. In addition, our methodology demonstrates full compatibility with both top-down and bottom-up fabrication approaches, providing a versatile platform for quantum engineering in oxides. 

\section{Results and Discussion}

\begin{figure}[ht]
    \centering
    \includegraphics[width=\linewidth]{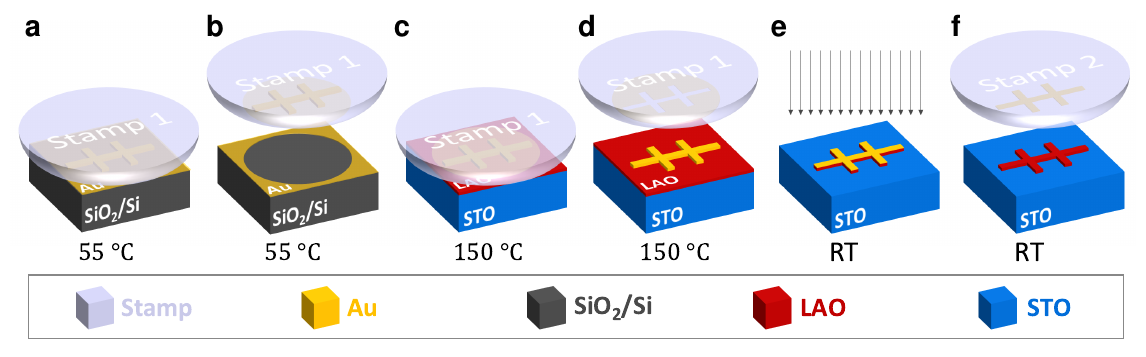}
    \caption{Illustration of an Au Hall bar mask transfer in the top-down patterning approach. (a,b) A stamp is used to pick up a pre-patterned Au Hall bar mask with the thin Au layer on $\mathrm{SiO_2}$/$\mathrm{Si}$. (c,d) The Au Hall bar, along with the thin Au layer directly beneath it, is released on the LAO/STO sample. The thin Au layer outside the patterned area remains adhered to the stamp, thereby preventing direct contact between the stamp and the sample. (e) Oxygen-enriched $\mathrm{Ar^+}$ ion milling removes the exposed LAO. (f) A second stamp is used to pick up the Au mask on the LAO/STO.}
    \label{fig:1}
\end{figure}

To eliminate residue during patterning, we developed a contamination-free patterning technique based on transferable Au films as hard masks and oxygen-enriched $\mathrm{Ar^+}$ ion milling. Figure \ref{fig:1} and \ref{fig:s1} illustrate the implementation of this method in the top-down approach. Prior to oxide patterning, Au masks were prepared by first depositing a 5--\SI{10}{nm} Au layer on the whole surface of clean $\mathrm{SiO_2}$/$\mathrm{Si}$ substrates, and followed by depositing 80--\SI{100}{nm} Au film in patterns defined by ultraviolet lithography (UVL) or electron-beam lithography (EBL). Then we use a dome-shaped polymer stamp (NOA 81, Norland optical adhesive) to pick up a thick Au pattern as well as the thin Au layer within the stamp's contact area (see Figure \ref{fig:1}a, \ref{fig:1}b) at $\SI{55}{\degreeCelsius}$. Subsequently, the Au pattern is released on LAO/STO at $\SI{150}{\degreeCelsius}$, with the thin Au layer outside the patterned area remaining on the stamp (Figure \ref{fig:1}c, \ref{fig:1}d, \ref{fig:s2}). Notably, this is a completely dry transfer process that eliminates the need for polymer melting or solvent immersion, as well as contact between the polymer and the rest of the sample, thanks to the Au isolation layer (Supporting Movies SM1, SM2). Next, $\mathrm{Ar^+}$ ion milling is performed in an $\mathrm{Ar}$/$\mathrm{O_2}$ gas mixture at a ratio of 1:5 (Figure \ref{fig:1}e). Finally, the Au mask is removed by another stamp at room temperature, yielding a patterned LAO/STO Hall bar (Figure \ref{fig:1}f).

In the fabrication of 2D material transistors and flexible electronics, multiple methods have been developed to achieve dry transfer printing of metal electrodes \cite{liu2018approaching,kong2020doping,went2019new,adfm.202002549,liu2020,liu2022graphene,wang2022general,yang2023highly,wu2023all,qi2023graphene,zhang2024reliable,he2025} and complex ultrathin films\cite{jeong2014,Shu_2024,acsnano.4c00564} on arbitrary
substrates. However, these approaches exhibit limitations in patterning sensitive oxide interfaces, partially due to residues from the polymer  stamps, e.g., poly(methyl methacrylate) (PMMA)\cite{liu2018approaching,adfm.202002549,kong2020doping,wang2022general,yang2023highly,wu2023all,qi2023graphene,zhang2024reliable,Shu_2024,acsnano.4c00564,he2025}, polymer polypropylene carbonate (PPC)\cite{went2019new}, polyvinyl alcohol (PVA)\cite{liu2022graphene,acsnano.4c00564} or polystyrene (PS)/poly(4-vinyl pyridine) (P4VP)\cite{jeong2014}. Further immersion in acetone, chloroform, water or other solvents is necessary to remove the residues. In addition, these methods typically require a pre-functionalization layer such as hexamethyldisilazane (HMDS)\cite{liu2018approaching,kong2020doping,went2019new,yang2023highly}, octadecyltrichlorosilane (OTS)\cite{Shu_2024} or a sacrificial graphene interlayer\cite{liu2022graphene,wang2022general,qi2023graphene} to reduce strong adhesion to silicon substrates, which complicates the experimental workflow and risks introducing additional contaminants to oxide surfaces. Here, we employ NOA 81 as the stamp material, which is optically clean and nearly residue-free. The stamp is prepared by depositing a droplet onto a glass slide to form a dome structure, followed by UV curing (see Supporting Information SI). This method enables efficient transfer of prefabricated Au patterns from $\mathrm{SiO_2}$/$\mathrm{Si}$ substrates to target samples. Remarkably, the whole transfer process takes less than \SI{10}{\min}, including the curing of the stamp.

Under the specified transfer condition, Au films spanning $5$--$\SI{100}{nm}$ thicknesses can be consistently picked up by the polymer stamp. Notably, the success rate of releasing exhibits strong thickness dependence, with thicker films showing significantly higher release rates than thinner films.
We attribute this behavior to the mechanical fragility of thin films: their tendency to form structural imperfections (e.g., wrinkles, microcracks) under stamp-induced deformation stress\cite{Niu2007Thickness,LU20101679} could substantially reduce interfacial adhesion to target substrates (Figure \ref{fig:s3}).

Conventional dry transfer printing strategies typically aim for damage-free transfer by minimizing in-plane stress to prevent film fracture \cite{ss.2025.52}. In contrast, our method deliberately leverages this inherent mechanical fragility. We exploit the strong thickness-dependent stability: thin (5--10 nm) Au layers are prone to wrinkling and cracking, while thick (80--100 nm) Au films remain robust under the same condition. This differential behavior enables a selective transfer process. The thin Au layer outside the patterned area remains adhered to the stamp, acting as an isolation layer to protect the oxide surface from direct contact with the polymer stamp. Crucially, the thin Au layer beneath the thick pattern is picked up together, forming a cohesive bilayer that is released successfully. This design effectively prevents the transfer of any contaminant from the stamp to the sample in the release step at elevated temperatures (see Figure \ref{fig:s4}), while simultaneously maintaining transfer success rates of greater than 90\% for thick Au films. Thus, by strategically controlling the location of interfacial adhesion failure, we achieve residue‑free patterning—transforming a common failure mode into an enabling feature of the fabrication process.

It is well-known that $\mathrm{Ar^+}$ ion bombardment can easily create oxygen vacancies in STO and induce metallicity\cite{reagor2005highly}.
To suppress the formation of oxygen vacancies, we explored $\mathrm{Ar^+}$ ion milling under oxygen-rich conditions. We found that an $\mathrm{Ar}$/$\mathrm{O_2}$ gas mixture ratio of 1:5 effectively etches STO without the formation of conductive channels that shunt devices.
By varying the chamber pressure and bias voltage, different etch rates can be achieved. For samples in this work, the etch rate is about \SI{2}{nm/s} for LAO or STO and 4--\SI{5}{nm/s} for transferred Au films.

\begin{figure}[ht]
    \centering
    \includegraphics[width=1\linewidth]{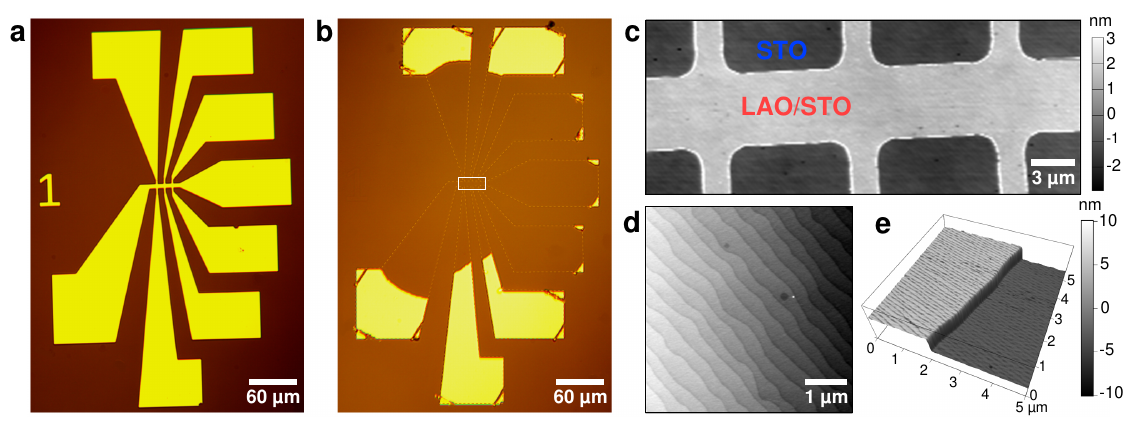}
    \caption{A Hall bar fabricated using a transferred Au mask in the top-down approach. (a) The transferred Au Hall bar pattern on LAO/STO. (b) The final LAO/STO Hall bar device. (c) AFM topography image of area highlighted within the white rectangle in (b). (d) AFM topography image of the surface of the Hall bar, preserving the atomic terraces of the pristine LAO/STO sample. (e) The abrupt edge between the device and the etched region.}
    \label{fig:2}
\end{figure}

Figure \ref{fig:2}a presents an optical micrograph of a transferred Au Hall bar structure ($\SI{20}{\mu m}\times\SI{5}{\mu m}$) with 8 electrodes on a (3+8)~{uc} LAO/STO sample\cite{nwaf156} (Supporting Movie SM1). The final device after mask removal is shown in Figure \ref{fig:2}b. 
By scratching the Au film with a tungsten tip at the corners and controlling the stamp’s contact area, a fraction of the mask is intentionally kept to serve as visual markers for wire bonding. This enables the subsequent precise bonding of aluminum wires to make ohmic contacts to the interface. The AFM topography image on the main channel of the Hall bar indicates successful pattern transfer (Figure \ref{fig:2}c).  Atomic terraces of the LAO/STO sample are clearly resolved (Figure \ref{fig:2}d), suggesting the pristine surface is preserved during this clean transfer process. Furthermore, optimized ion milling yields well-defined edges with a sharp transition between the device and the etched region (Figure \ref{fig:2}e). Systematic AFM characterization performed on dozens of fabricated LAO/STO devices confirms the residue-free nature of our fabrication process. The absence of any observable contamination is consistently observed, as exemplified by the topography across both the device channel and the etched region in the representative device shown in Figures {\ref{fig:2}c} and {\ref{fig:2}e}.

\begin{figure}[ht]
    \centering
    \includegraphics[width=1\linewidth]{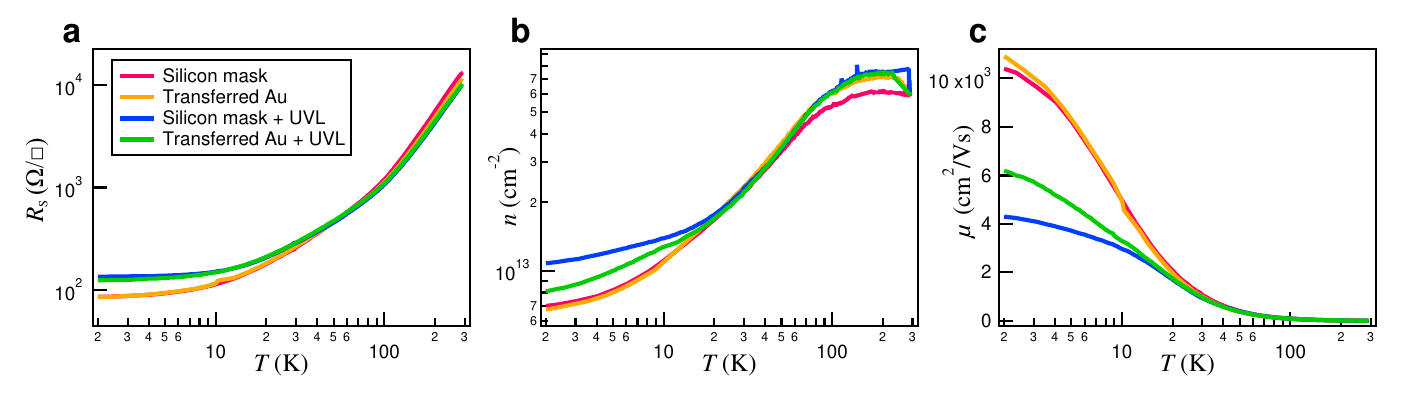}
    \caption{Temperature-dependent transport properties of patterned squares on Sample A: (a) sheet resistance $R_s$, (b) carrier density $n$, and (c) mobility $\mu$ as functions of temperature extracted from vdP measurement.}
    \label{fig:3}
\end{figure}

To demonstrate that the pristine interface is preserved electronically, we conducted a comparative transport study of devices fabricated with silicon mask, transferred Au mask, and UVL methods. Initially, a $\SI{1}{mm}\times\SI{1}{mm}$ square defined by a silicon stencil mask and a $\SI{200}{\mu m}\times\SI{200}{\mu m}$ square defined by a transferred Au mask were fabricated on the same $\SI{5}{mm}\times\SI{5}{mm}$ high mobility (3+2)~{uc} LAO/STO sample (Sample A). After fast van der Pauw (vdP) measurement\cite{nwaf156} at cryogenic temperature, UVL was employed to pattern a $\SI{200}{\mu m}\times\SI{200}{\mu m}$ square within the silicon-masked region. Meanwhile, the area fabricated by the Au mask method was only subjected to photoresist (PR) coating and removal during the same process. Finally, vdP measurements were performed on both regions again. As shown in Figure \ref{fig:3}, the temperature-dependent characteristics of mobility and carrier density of the Au-masked square closely resemble those of the silicon-masked square, with very high mobilities exceeding $\SI{10000}{cm^2/Vs}$ at  \SI{2}{\kelvin}. In contrast, the sheet resistances and carrier densities are significantly higher in the devices subjected to PR processing, suggesting more carriers are created due to the formation of oxygen vacancies. Indeed, the measured mobilities dropped to $\sim\SI{5000}{cm^2/Vs}$, suggesting severe degradation of the interfaces. The direct comparison between pre- and post-UVL process of the Au-masked region conclusively demonstrates that the conventional UVL process causes interface deterioration, ruling out the possibility of regional inhomogeneity. To exclude potential temporal degradation, we have conducted repeated measurements over several days before UVL, confirming the stability of the interface properties.

\begin{figure}[ht]
    \centering
    \includegraphics[width=\linewidth]{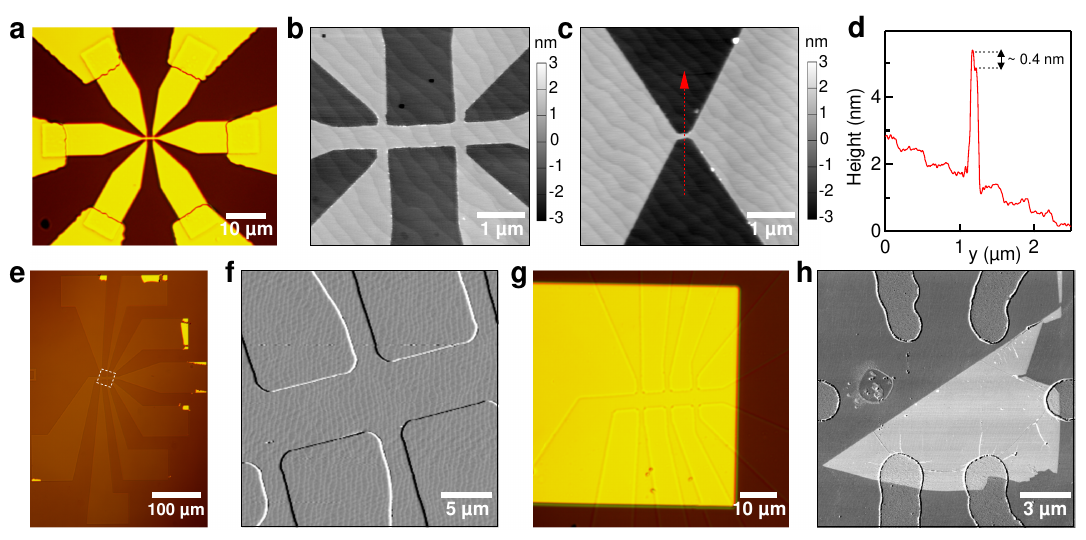}
    \caption{Versatile device fabrication. (a) The two-step-transferred nanoscale Au Hall bar pattern. (b) The final LAO/STO Hall bar device. (c) An LAO/STO nanoconstriction with a width $\sim$\SI{100}{nm}. (d) The height profile along the red dashed line in (c). (e) A Hall bar fabricated via the bottom-up approach. (f) AFM topography image of the area highlighted within the white dashed square in (e). (g) A Hall bar device with an Au top gate. (h) A graphene monolayer transferred onto a STO substrate with pre-patterned electrodes.}
    \label{fig:4}
\end{figure}

Notably, our method successfully extends to nanoscale device fabrication through a two-step transfer process. We first transferred a nanoscale Au Hall bar defined by EBL and then transferred UVL-patterned Au contact pads. The Au pads were carefully aligned under a microscope and laminated onto the first layer with controlled overlap (Figure \ref{fig:4}a and Supporting Movie SM3). Figure \ref{fig:4}b shows a $\SI{1.6}{\mu m}\times\SI{0.4}{\mu m}$ LAO/STO Hall bar structure with four \SI{200}{nm} wide voltage leads. Note that the atomic terraces are well-preserved across the entire device, including the insulating regions. Figure \ref{fig:4}c shows a functional \SI{100}{nm} wide LAO/STO nanoconstriction. A clearly resolved terrace ($\sim$\SI{0.4}{nm}) crossing the channel demonstrates the preservation of the original surface by the Au mask (Figure \ref{fig:4}d). It is noteworthy that channels narrower than \SI{500}{nm} exhibit increased susceptibility to deformation and cracking during transfer, thus requiring the use of flatter stamps, thicker Au and smaller contact areas to minimize mechanical strain. With this careful parameter optimization, our method proves to be robust and reproducible, as demonstrated by the consistent results across multiple transfers of patterns with nanoscale features (Supporting Movie SM 4). In contrast, the photoresist in the conventional EBL process exhibits poor resistance to oxygen-enriched $\mathrm{Ar^+}$ ion milling, consequently resulting in failed pattern transfer.

The technique of transferring Au masks can be adapted to the bottom-up approach (Figure \ref{fig:s6}). Figure \ref{fig:4}e,f show a Hall bar fabricated by selectively depositing LAO on STO. The fabrication began with transferring a predefined Au Hall bar pattern onto the STO substrate, followed by etching the exposed regions to a depth of \SI{10}{nm}. To avoid the formation of electron gas outside the Hall bar region, we deposited a \SI{10}{nm} $\mathrm{SiO_2}$ layer. Subsequently, the Au pattern with $\mathrm{SiO_2}$ covered was picked up to expose only the designated Hall bar region. The final conductive Hall bar was then formed through controlled deposition of several LAO unit cells using pulsed laser deposition (PLD). Apart from patterning, this technique offers significant advantages for device integration by enabling direct transfer of top gates onto pre-fabricated structures without any additional lithography steps. This capability is illustrated in Figure \ref{fig:4}g, which shows a Hall bar device covered by a rectangular Au top gate electrode. Furthermore, two-dimensional materials can also be transferred directly with this stamp (Figure \ref{fig:4}h).

It is noteworthy that, two-dimensional materials such as graphene \cite{son2018} and $\mathrm{CrOCl}$ \cite{venkatram2026} have recently been explored as masks for lithography, leveraging their etch selectivity at specific etching conditions. Similarly, while both strategies employ transferred thin films as masks to protect the underlying surface, our approach using Au films provides a more versatile and scalable platform. Compared to the exfoliation and pre-patterning of two-dimensional materials, Au masks allow for standard fabrication via conventional lithography and evaporation, offering higher design freedom and significantly lower process complexity. This suggests the significant potential of Au masks for the future high-throughput fabrication of sensitive oxide electronic devices, while consistently maintaining a high level of quality.

\section{Conclusion}

In summary, we have developed a versatile patterning method based on transferred Au masks and oxygen-enriched $\mathrm{Ar^+}$ ion milling. The residue-free transfer process, coupled with the protection of Au layers, effectively preserves the pristine quality of the oxide interfaces by circumventing the chemical and polymeric contamination inherent in conventional methods. This approach provides a powerful platform for probing intrinsic properties of as-grown interfaces at the nanoscale, offering new insights into the conductive behaviors and complex electron correlations at oxide heterostructures. Furthermore, the method demonstrates broad applicability beyond interface studies, including lithography-free gate integration and dry transfer of two-dimensional materials.

\section{Experimental Section}

\subsection{Au Mask Fabrication}

The transferable Au masks are fabricated through the following procedure. First, the 5–10 nm thin Au film is deposited on a clean $\mathrm{SiO_2}$(\SI{300}{nm})/$\mathrm{Si}$ substrate via electron-beam evaporation at a rate of 0.5--\SI{0.8}{\text{Å}/s}. Subsequently, the 80--\SI{100}{nm} thick Au patterns for transfer are fabricated on this Au-coated substrate. The patterns are defined by either UVL or EBL, followed by Au deposition via electron-beam evaporation at a rate of 1--\SI{2}{\text{Å}/s}.

\subsection{Stamp Preparation}

The polymer stamp is prepared using the optical adhesive NOA 81. A droplet of NOA 81 is deposited onto a clean glass slide to form a dome-shaped structure with a diameter of approximately \SI{3}{mm}. The adhesive is then cured for \SI{3}{min} under a \SI{365}{nm} ultraviolet lamp with a power density of \SI{1}{V/cm^2}, resulting in a solid, optically transparent stamp ready for the dry transfer process.

\subsection{Au Mask Transfer}

The transfer procedure for thick Au patterns fabricated on both pristine and Au-coated $\mathrm{SiO_2}$/$\mathrm{Si}$ substrates is identical. The glass slide bearing the cured polymer stamp is inverted and mounted onto a micromanipulator. Separately, the $\mathrm{SiO_2}$/$\mathrm{Si}$ substrate with the pre-fabricated Au patterns is fixed to another micromanipulator stage and positioned directly beneath the stamp. The stage holding the substrate is first heated to \SI{55}{\celsius}. The stamp is then lowered until it makes full contact with the target Au pattern. After maintaining contact for \SI{2}{min}, the stamp is lifted rapidly, picking up the Au pattern. Subsequently, the target oxide sample (e.g., LAO/STO) is placed on the heating stage beneath the stamp. The stage temperature is raised to \SI{150}{\celsius}. The stamp, now carrying the Au pattern, is carefully lowered until the pattern makes complete contact with the oxide surface. After 30--\SI{60}{s}, the stamp is lifted slowly, releasing the thick Au pattern onto the target surface.

\subsection{Oxygen-enriched $\mathrm{Ar^+}$ Ion Milling}

The oxygen-enriched $\mathrm{Ar^+}$ ion milling is performed using either a custom-built ion milling system or a commercial system (Oxford PlasmaPro NGP80). When using the Oxford PlasmaPro NGP80, the etching is conducted with a gas flow of \SI{10}{sccm} $\mathrm{Ar}$ and \SI{50}{sccm} $\mathrm{O_2}$, at a radio-frequency (RF) power of \SI{150}{W}. The chamber pressure varies between \SI{20}{mTorr} and \SI{50}{mTorr}, resulting in different etch rates.


\section*{Acknowledgements}

This work was supported by the National Key Research and Development Program of China (2024YFA1409500) and the CAS Project for Young Scientists in Basic Research (YSBR-100). This work was partially carried out at the USTC Center for Micro and Nanoscale Research and Fabrication.


\section*{Supporting Information}

\begin{itemize}
  \item Front-view schematic of the fabrication process in Figure 1 (Figure \ref{fig:s1}); stamps after the transfer process (Figure \ref{fig:s2}); thickness-dependent Au pattern transfer behavior (Figure \ref{fig:s3}); AFM characterization of the oxide surface before and after the Au mask transfer process (Figure \ref{fig:s4}); temperature-dependent transport properties of patterned squares on a low-mobility LAO/STO sample (Figure \ref{fig:s5}); comparison of the bottom-up and the top-down patterning approaches (Figure \ref{fig:s6}). (PDF)
  \item Movie SM1: Transfer process of the Au Hall bar in Figure \ref{fig:2}a. (MP4)
  \item Movie SM2: Transfer process of a 6-pad Au Hall bar. (MP4)
  \item Movie SM3: Transfer process of the contact pads in Figure \ref{fig:4}a. (MP4)
  \item Movie SM4: Sequential successful transfers of multiple EBL-defined nanoscale patterns. (MP4)
\end{itemize}


\bibliography{manuscript}

\setcounter{figure}{0}
\renewcommand{\thefigure}{S\arabic{figure}}

\begin{figure}[p]
    \centering
    \includegraphics[width=\linewidth]{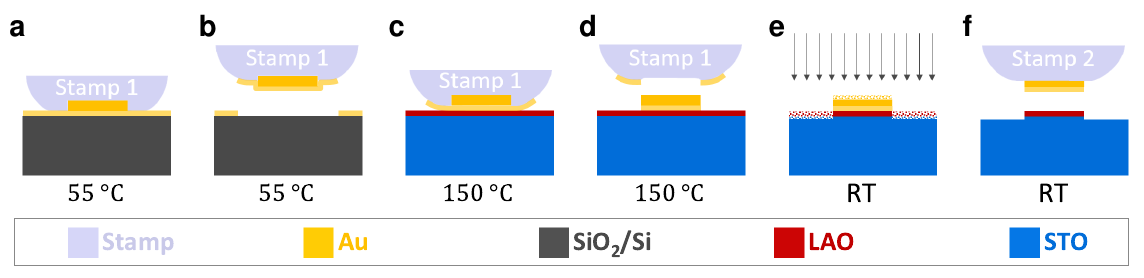}
    \caption{The front-view schematic of the fabrication process in Figure 1. (a,b) A stamp is used to pick up a pre-patterned Au Hall bar mask with a Au isolation layer on $\mathrm{SiO_2}$/$\mathrm{Si}$. (c,d) The Au Hall bar, along with the thin Au layer directly beneath it, is released on the LAO/STO sample. The Au isolation layer outside the patterned area remains adhered to the stamp, thereby preventing direct contact between the stamp and the sample. (e) Oxygen-enriched $\mathrm{Ar^+}$ ion milling removes the exposed LAO. (f) A second stamp is used to pick up the Au mask on the LAO/STO.}
    \label{fig:s1}
\end{figure}
\clearpage

\begin{figure}[p]
    \centering
    \includegraphics[width=\linewidth]{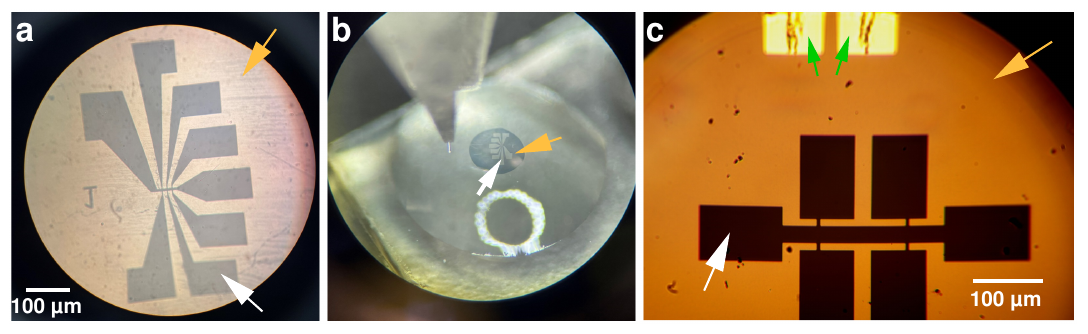}
    \caption{Stamps after the transfer process. The thin Au films outside the patterned areas remain on the stamps (yellow arrows), while the thick Au patterns have been transferred onto target oxide surfaces, exposing the underlying stamp surfaces (white arrows). The green arrows indicate parts of a neighboring thick Au pattern. (a)(b) The stamp shown in Supporting Movie SM2. (c) The stamp shown in Supporting Movie SM3. The particles were introduced accidentally during the process of removing the stamp from the glass slide.}
    \label{fig:s2}
\end{figure}
\clearpage

\begin{figure}[p]
    \centering
    \includegraphics[width=\linewidth]{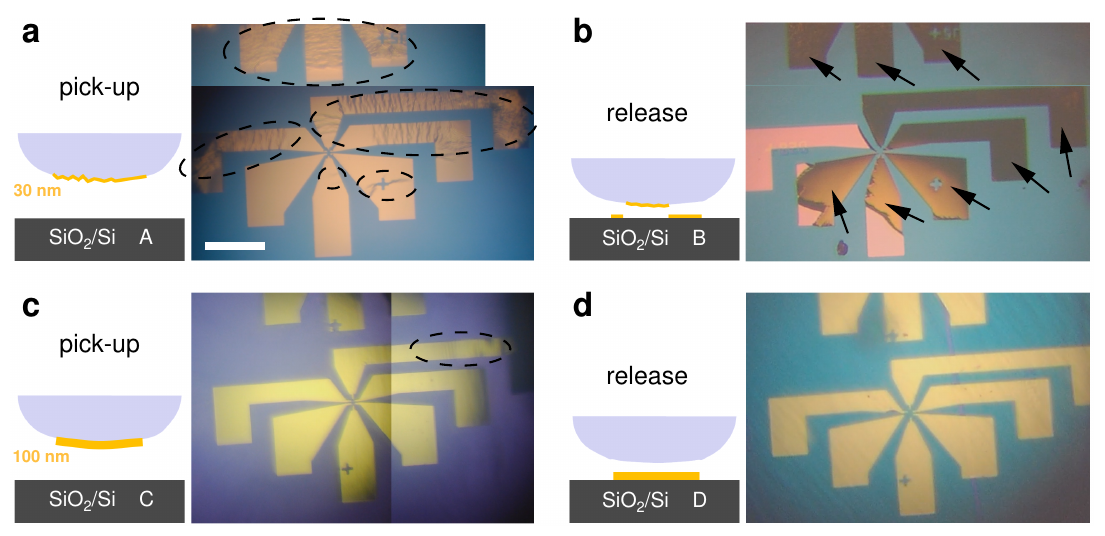}
    \caption{Thickness-dependent Au pattern transfer behavior. Schematic illustrations (left) and corresponding top-view optical micrographs through the stamp (right) are shown for each case. (a) After picking up a 30-nm-thick Au pattern (without the thin Au isolation layer), extensive wrinkles and micro-cracks are observed (black dashed circle). (b) After the release attempt, only portions of the 30-nm-thick pattern are successfully transferred onto $\mathrm{SiO_2}$/$\mathrm{Si}$. Arrows indicate regions that remain adhered to the stamp. (c) After being picked up, the 100-nm-thick Au pattern (without the isolation layer) remains smooth and crack-free, exhibiting only minor undulations. (d) The entire 100-nm-thick pattern is successfully released onto $\mathrm{SiO_2}$/$\mathrm{Si}$. Scale bar: {\SI{100}{\mu m}} (applies to all optical micrographs). The optical micrographs in panels (a)(b)(c) are composites of two image frames stitched together due to the limited field of view.}
    \label{fig:s3}
\end{figure}
\clearpage

\begin{figure}[p]
    \centering
    \includegraphics[width=\linewidth]{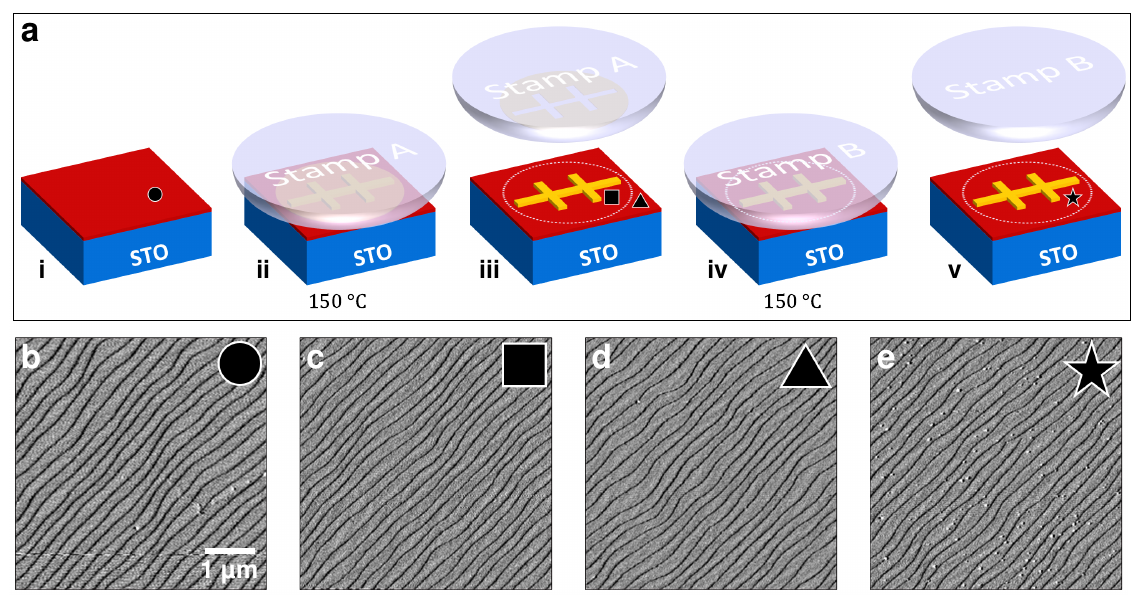}
    \caption{AFM characterization of the oxide surface before and after the Au mask transfer process. (a) Illustration of the treatments on the surface. (b) The original surface was clean before the treatments. (c) The region protected by an Au layer is still clean. (d) The region at the periphery of the stamp's contact area is still clean. (e) The region that was contacted by a stamp at \SI{150}{\celsius} shows a few residues with $\sim$\SI{100}{nm} diameter and $\sim$\SI{1}{nm} height. In contrast, no such contamination is observed in the etched regions that were contacted by a stamp at room temperature, as shown in Figures 2c and 2e.}
    \label{fig:s4}
\end{figure}\clearpage

\begin{figure}[p]
    \centering
    \includegraphics[width=\linewidth]{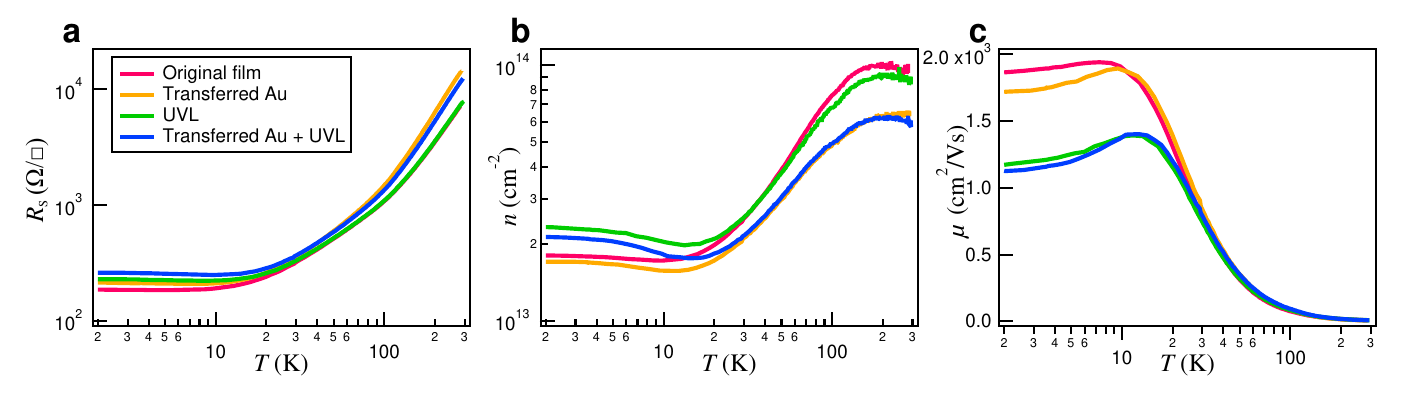}
    \caption{Temperature-dependent transport properties of patterned squares on a low-mobility LAO/STO sample: (a) sheet resistance $R_s$, (b) carrier density $n$, and (c) mobility $\mu$ as functions of temperature (T = 2--\SI{300}{K}) extracted from vdP measurements. The minor mobility difference ($\sim$ 8\%) between the original film and the region patterned with the transferred Au mask is attributed to the intrinsic spatial inhomogeneity commonly observed in low-mobility LAO/STO interfaces. This variation is notably smaller than the substantial degradation caused by the UVL process.
    }
    \label{fig:s5}
\end{figure}\clearpage

\begin{figure}[p]
    \centering
    \includegraphics[width=\linewidth]{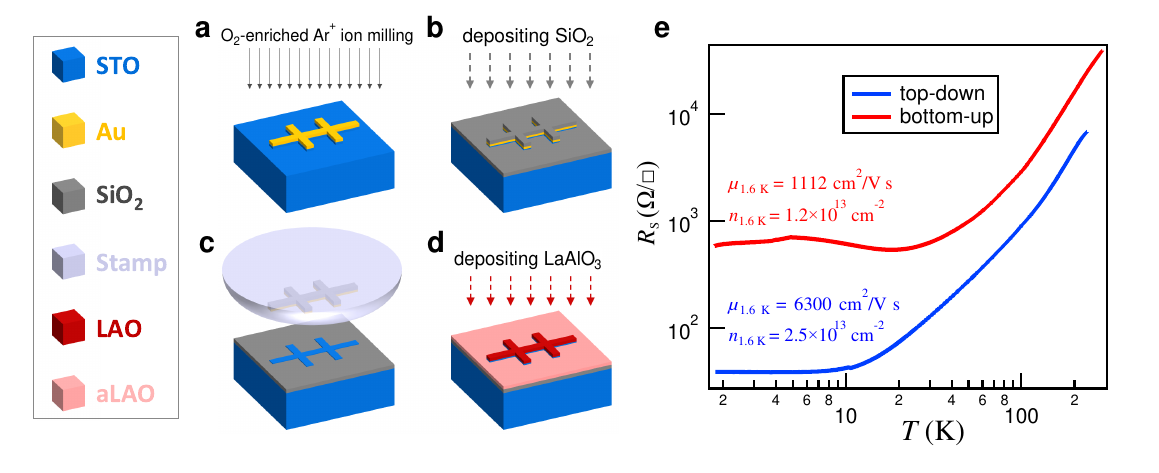}
    \caption{Comparison of the bottom-up and the top-down patterning approaches. (a--d) Illustration of the bottom-up patterning approach: (a) Perform oxygen-enriched $\mathrm{Ar^+}$ ion milling on an STO substrate with a pre-transferred Au Hall bar pattern. (b) Deposit $\mathrm{SiO_2}$ via electron-beam evaporation. (c) Remove the Au pattern along with the overlying $\mathrm{SiO_2}$ using a stamp. (d) Depositing LAO by PLD. The areas outside the Hall bar, being covered by $\mathrm{SiO_2}$, form amorphous LAO. (e) Comparison of temperature-dependent sheet resistance $R_s$ of Hall bars patterned by the top-down approach and the bottom-up approach. The LAO deposition parameters (e.g., laser energy, pressure, temperature) are nominally identical. The differences in electronic properties are attributed to the inherent limitation of the bottom-up approach, where the $\mathrm{SiO_2}$ obstructs the substrate area, preventing effective in-situ growth monitoring by reflection high-energy electron diffraction (RHEED) and potentially leading to variations in the interfacial properties compared to the uniformly grown film used in the top-down approach. }
    \label{fig:s6}
\end{figure}

\end{document}